\pdfoutput=1
\RequirePackage{ifpdf}
\ifpdf
\documentclass[pdftex]{sigma}
\else
\documentclass{sigma}
\fi

\newcommand{\mysub}[1]{_{\scriptscriptstyle #1}}
\renewcommand\vec[1]{\boldsymbol{#1}}
\renewcommand\Re{\mathop{\mbox{Re}}}
\renewcommand\Im{\mathop{\mbox{Im}}}
\newcommand\mypmatrix[1]{\begin{pmatrix}#1\end{pmatrix}}

\newcommand{\tdhlT}[1]{\mathbb{T}\mysub{#1}}
\newcommand{\tdhlshifted}[2]{\tdhlT{#1}#2}
\newcommand{\tdhlShifted}[2]{\left(\tdhlT{#1}#2\right)}

\newcommand{\tdhlz}[1]{%
  \ifcase #1 \mathfrak{a}
  \or \mathfrak{b}^{(1)}
  \or \mathfrak{b}^{(2)}
  \or \mathfrak{c}^{(1)}
  \or \mathfrak{c}^{(2)}
  \or \mathfrak{b}^{(1,2)}
  \or \mathfrak{c}^{(1,2)}
  \else \stackrel{#1}{\mathfrak{o}}\fi}

\newcommand{\tdhlZ}[1]{%
  \ifcase #1 ?
  \or \mathfrak{Q}
  \or \mathfrak{R}
  \or \mathfrak{P}
  \else ?\fi}

\newcommand{\tdhlMatrix}[2]{\mathsf{#1}_{#2}}
\newcommand{\tdhlBra}[2]{\langle{\mathsf{#1}}_{#2}|}

\newcommand{\tdhlmatrix}[2][]{\mathsf{#2}\mysub{#1}}
\newcommand{\tdhlbra}[2][]{\langle{\mathsf{#2}}\mysub{#1}|}
\newcommand{\tdhlket}[1]{|{#1}\rangle}
\newcommand{\tdhlL}{\tdhlmatrix{L}}
\newcommand{\tdhlR}{\tdhlmatrix{R}}

\begin{document}

\numberwithin{equation}{section}

\newtheorem{prop}{Proposition} \numberwithin{prop}{section}

\allowdisplaybreaks

\renewcommand{\PaperNumber}{044}

\FirstPageHeading

\ShortArticleName{Two-Dimensional Toda--Heisenberg Lattice}

\ArticleName{Two-Dimensional Toda--Heisenberg Lattice}

\Author{Vadim E.~VEKSLERCHIK}
\AuthorNameForHeading{V.E.~Vekslerchik}

\Address{Institute for Radiophysics and Electronics of NAS of Ukraine,
\\
12, Proskura Str., Kharkiv, 61085, Ukraine}
\Email{\href{mailto:vekslerchik@yahoo.com}{vekslerchik@yahoo.com}}

\ArticleDates{Received February 06, 2013, in f\/inal form June 04, 2013; Published online June 12, 2013}

\Abstract{We consider a~nonlinear model that is a~combination of the anisotropic two-dimensional classical
Heisenberg and Toda-like lattices.
In the framework of the Hirota direct approach, we present the f\/ield equations of this model as
a~bilinear system, which is closely related to the Ablowitz--Ladik hierarchy, and derive its $N$-soliton
solutions.}


\Classification{39A14; 82D40; 35C08; 11C20}

\section{Introduction}

In this paper we consider a~two-dimensional lattice that can be viewed as a~generalization of the
anisotropic two-dimensional classical Heisenberg model~\cite{BL55,M81},
\begin{gather*}
\mathcal{E}_{\mathrm{Heis}}=\sum_{\vec{r}\in\Lambda}\sum_{i=1,2}J_{i}\left(\vec{\phi}_{\vec{r}},\vec{\phi}
_{\vec{r}+\vec{\delta}_{i}}\right),
\end{gather*}
where $\vec{\phi}_{\vec{r}}$ is a~three-dimensional unit vector,
\begin{gather*}
\left(\vec{\phi}_{\vec{r}},\vec{\phi}_{\vec{r}}\right)=1
\end{gather*}
(with brackets standing for the standard scalar product), $J_{1,2}$ are the constants characterizing the
interaction between near-neighbour sites (exchange constants)                
and $\Lambda$ is a~two-dimensional lattice
formed by two vectors $\vec{\delta}_{1}$ and $\vec{\delta}_{2}$:
\begin{gather*}
\Lambda=\left\{m_{1}\vec{\delta}_{1}+m_{2}\vec{\delta}_{2}\right\}_{m_{1},m_{2}=0,\pm1,\pm2,\dots}.
\end{gather*}
The generalization that we are going to study consists in replacing the \emph{constants} $J_{i}$ with some
functions of new variables.
In more details, we associate with each site, in addition to the vector~$\vec{\phi}_{\vec{r}}$, a~new
variable $u_{\vec{r}}$ and modify the exchange constants as
\begin{gather}
J_{i}\to J_{i}\exp\left\{u_{\vec{r}}-u_{\vec{r}+\vec{\delta}_{i}}\right\}.
\label{def-u}
\end{gather}
An elementary example that leads to the above modif\/ication of the exchange interaction is to permit the
spins to oscillate in the direction perpendicular to the plane and to state that the interaction
coef\/f\/icients $J$ depend on the distance (in the three-dimensional space) between the spins:
$\mathcal{E}_{ab}=J\left(|\vec{R}_{a}-\vec{R}_{b}|\right)\left(\vec{\phi}_{\vec{r}_{a}},\vec{\phi}_{\vec{r}_{b}}\right)$, where $\vec{R}_{a}=\vec{r}_{a}+u_{\vec{r}_{a}}\vec{\nu}$, with $\vec{\nu}\bot\vec{\delta}_{1,2}$.
In this case the interaction between the nearest neighbours depends on $\left|\vec{\delta}_{i}\right|$ and
$\left|u_{\vec{r}}-u_{\vec{r}+\vec{\delta}_{i}}\right|$, that can be modeled by~\eqref{def-u}.
Of course, the dependence given by \eqref{def-u} is far from being realistic, however this toy model can
give some insight into ef\/fects caused by such kind on nonlinearities, and especially into the possibility
of appearing of specif\/ic structures like solitons that are discussed in this paper.

To summarize, our model is described by the energy functional
\begin{gather}
\mathcal{E}=\sum_{\vec{r}\in\Lambda}\sum_{i=1,2}J_{i}\exp\left\{u_{\vec{r}}-u_{\vec{r}+\vec{\delta}_{i}}
\right\}\left(\vec{\phi}_{\vec{r}},\vec{\phi}_{\vec{r}+\vec{\delta}_{i}}\right).
\label{energy-vec}
\end{gather}
It is easy to see that neglecting the $\vec{\phi}_{\vec{r}}$-part, or imposing the restrictions
$\left(\vec{\phi}_{\vec{r}},\vec{\phi}_{\vec{r}+\vec{\delta}_{i}}\right)=c_{i}$ for all~$\vec{r}$ and
redef\/ining the constants $J_{i}$, one arrives at the one of the Hirota's versions of the discrete 2D Toda
lattice~\cite{H77a,H77b,PGR95},
\begin{gather}
\mathcal{E}_{\mathrm{Toda}}=\sum_{\vec{r}\in\Lambda}\sum_{i=1,2}J_{i}\exp\left\{u_{\vec{r}}-u_{\vec{r}
+\vec{\delta}_{i}}\right\},
\label{energy-Toda}
\end{gather}
whose f\/ield equations $\delta\mathcal{E}_{\mathrm{Toda}}/\delta u_{\vec{r}}=0$ are known to be integrable.
Thus, we call model \eqref{energy-vec}, which is the subject of this paper, the two-dimensional
Toda--Heisenberg lattice (2DTHL).

\begin{figure}[t]
\centering
\includegraphics[scale=1]{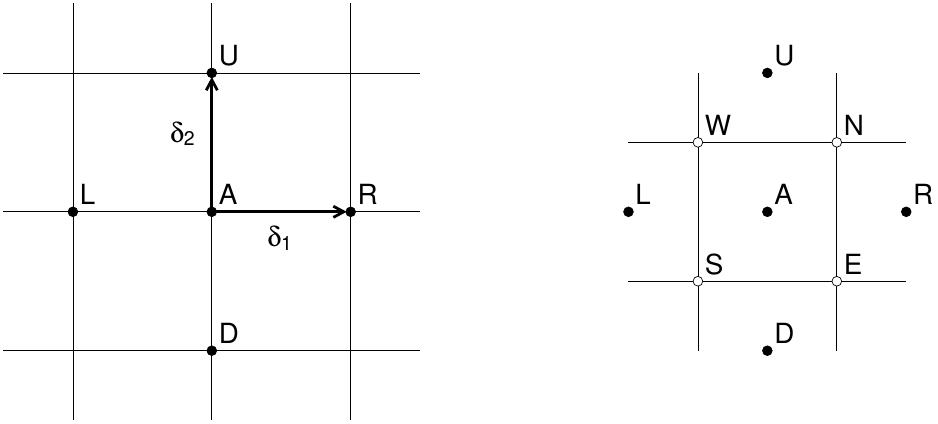}

\caption{Direct and dual lattices, nearest-neighbour notation.}
\label{fig-lattice}
\end{figure}

To make the following formulae more readable we introduce the alternative notation: instead of the vector
index we will use a~letter one,
\begin{gather*}
\vec{\phi}_{\vec{r}},u_{\vec{r}}\to\vec{\phi}\mysub{A},u\mysub{A},
\end{gather*}
and denote the nearest neighbours of the point $A$ as indicated in Fig.~\ref{fig-lattice} ($R$, $L$,
$U$ and $D$ stand for `right', `left', `up', `down').
The energy of the model can be rewritten as
\begin{gather}
\mathcal{E}=\frac{1}{2}\sum\mysub{A}\mathcal{E}\mysub{A},
\label{energy-let}
\end{gather}
where
\begin{gather}
\mathcal{E}\mysub{A}=\sum\mysub{B}J\mysub{B}\,\exp\left\{\varepsilon\mysub{B}\left(u\mysub{A}-u\mysub{B}
\right)\right\}\left(\vec{\phi}\mysub{A},\vec{\phi}\mysub{B}\right)
\label{energy-a}
\end{gather}
and the summation index runs over the nearest neighbours,
\begin{gather*}
\sum\mysub{B}\cdots=\sum\mysub{B=R,U,L,D}\cdots.
\end{gather*}
The constants $J\mysub{B}$ are nothing but $J_{1,2}$,
\begin{gather*}
J\mysub{R}=J\mysub{L}=J_{1},
\qquad
J\mysub{U}=J\mysub{D}=J_{2},
\end{gather*}
while $\varepsilon\mysub{B}$ takes
into account the signs of $u\mysub{B}$ in the arguments of the
exponential functions,
\begin{gather*}
\varepsilon\mysub{R}=-\varepsilon\mysub{L}=\varepsilon\mysub{U}=-\varepsilon\mysub{D}=1
\end{gather*}
(see Table~\ref{notation}).
The central object of the presented study are the Euler--Lagrange equations corresponding to
\eqref{energy-let} with \eqref{energy-a},
\begin{gather}
\frac{\delta\mathcal{E}\mysub{A}}{\delta u\mysub{A}}=\sum\mysub{B}\varepsilon\mysub{B}J\mysub{B}
\exp\{\varepsilon\mysub{B}\left(u\mysub{A}-u\mysub{B}\right)\}\left(\vec{\phi}\mysub{A},\vec{\phi}\mysub{B}
\right)=0,
\label{euler-u}
\\
\frac{\delta\mathcal{E}\mysub{A}}{\delta\vec{\phi}\mysub{A}}=\sum\mysub{B}J\mysub{B}
\exp\{\varepsilon\mysub{B}\left(u\mysub{A}-u\mysub{B}\right)\}\left[\vec{\phi}\mysub{A}\times\vec{\phi}
\mysub{B}\right]=0.
\label{euler-phi}
\end{gather}
Here, we have calculated the derivative with respect to $\vec{\phi}\mysub{A}$ bearing in mind the fact that
$\left|\vec{\phi}\mysub{A}\right|=1$, which implies that the admissible variations should be orthogonal to
$\vec{\phi}\mysub{A}$, which in its turn leads to the `def\/inition'
$\delta\left(\vec{\phi}\mysub{A},\vec{\psi}\right)/\delta\vec{\phi}\mysub{A}=\left[\vec{\phi}\mysub{A},\vec{\psi}\right]$.
The same result, \eqref{euler-phi}, can be reproduced by means of the Lagrange multipliers.
\begin{table}[t]
\centering
\caption{Nearest-neighbour notation and parameters.}
\label{notation}
\vspace{1mm}
\begin{tabular}
{|c|c|c|c|c|c|c|}
\hline
\strut $B$ & $B'$ & $B''$ & $J_{B}$ & $\varepsilon_{B}$ & $\xi_{B}$ & $\mu_{B}$
\\
\hline
$R$&$E$&$N$&$J_{1}$&$+1$&$\xi_{1}$&$\mu$
\\
$U$&$N$&$W$&$J_{2}$&$+1$&$\xi_{2}$&$-1/\mu$
\\
$L$&$W$&$S$&$J_{1}$&$-1$&$\xi_{1}$&$\mu$
\\
$D$&$S$&$E$&$J_{2}$&$-1$&$\xi_{2}$&$-1/\mu$
\\
\hline
\end{tabular}
\end{table}

\section{Bilinearization of the 2DTHL}

\looseness=-1
In this section we bilinearize the f\/ield equations \eqref{euler-u} and \eqref{euler-phi}.
This will be done in several steps.
First we replace the vector variables $\vec{\phi}\mysub{A}$ with scalar ones using a~parametrization which
can be viewed as an alternative to the stereographic projection.
Secondly, we reduce some of the f\/ive-site equations (star-equations) with simpler ones (quad-equations).
Then, we introduce the tau-functions and, f\/inally, split (in the next section) the obtained bilinear
equations into the standard three-term Hirota-like ones, which are closely related to the Ablowitz--Ladik
hierarchy (ALH).

\subsection{Scalar equations}

It is easy to check that any three-dimensional unit vector $\vec{\phi}$, $\vec{\phi}^{2}=1$, can be
presented in terms of a single 
complex function $q$ as
\begin{gather}
\vec{\phi}=\frac{1}{\sqrt{p}}\left(
\begin{matrix}
\Re q
\\
\Im q
\\
1
\end{matrix}
\right),
\label{vec-scal}
\end{gather}
where
\begin{gather}
p=1+|q|^{2}.
\label{def-p}
\end{gather}
In terms of $q$ the scalar and vector products are given by
\begin{gather*}
\left(\vec{\phi}\mysub{A},\vec{\phi}\mysub{B}\right)=\frac{1}{\sqrt{p\mysub{A}p\mysub{B}}}
\left(1+\Re q\mysub{A}q\mysub{B}^{*}\right),
\\
\left[\vec{\phi}\mysub{A}\times\vec{\phi}\mysub{B}\right]=\frac{1}{\sqrt{p\mysub{A}p\mysub{B}}}
\Im\left(q\mysub{A}-q\mysub{B}\right)\left(
\begin{matrix}
1
\\
-i
\\
-q\mysub{A}^{*}
\end{matrix}
\right)
\end{gather*}
(the asterisk denotes the complex conjugation).
Using these formulae and replacing $q\mysub{A}^{*}$ with the additional variable $r\mysub{A}$,
\begin{gather*}
r\mysub{A}=-q\mysub{A}^{*}
\end{gather*}
one can rewrite the f\/ield equations \eqref{euler-u} and \eqref{euler-phi} as follows
\begin{gather}
0=\sum\mysub{B}f\mysub{AB}\left(q\mysub{A}-q\mysub{B}\right),
\label{eq-q-a}
\\
0=\sum\mysub{B}f\mysub{AB}\left(r\mysub{A}-r\mysub{B}\right),
\label{eq-r-a}
\\
0=\sum\mysub{B}\varepsilon\mysub{B}f\mysub{AB}\left[1-{\tfrac{1}{2}}\left(q\mysub{A}r\mysub{B}
+q\mysub{B}r\mysub{A}\right)\right],
\label{eq-p-a}
\end{gather}
where
\begin{gather*}
f\mysub{AB}=J\mysub{B}\frac{\exp\left\{\varepsilon\mysub{B}\left(u\mysub{A}-u\mysub{B}\right)\right\}}
{\sqrt{p\mysub{A}p\mysub{B}}}.
\end{gather*}

\subsection{Quad-equations}

Now we arrive at the key moment of bilinearization of our equations.
It consists in introducing the dual lattice, as is shown in Fig.~\ref{fig-lattice}, whose nodes closest
to the point $A$ are denoted by the letters $N$, $W$, $S$ and $E$ (coming from `north', `west', `south',
and `east'),
and extending the functions $q\mysub{A}$ and $r\mysub{A}$ to the points of the dual lattice by
\begin{gather}
f\mysub{AB}\left(q\mysub{A}-q\mysub{B}\right)=c\left(q\mysub{B'}-q\mysub{B''}\right),
\label{ansatz-Q}
\\
f\mysub{AB}\left(r\mysub{A}-r\mysub{B}\right)=c\left(r\mysub{B'}-r\mysub{B''}\right),
\label{ansatz-R}
\end{gather}
where $B'$ and $B''$ are the ends of the oriented edge of the dual lattice that crosses the edge $(AB)$ of
the direct one (see Table~\ref{notation}) and $c$ is a~constant.
Of course, equations \eqref{ansatz-Q}, \eqref{ansatz-R} cannot be viewed as def\/initions of $q\mysub{B'}$
and $r\mysub{B'}$ because the determinants of the right-hand sides is zero.
We propose them as just an \emph{ansatz}.
Its role is that it `solves' 4-star equations \eqref{eq-q-a} and \eqref{eq-r-a}.
Indeed, the right-hand sides of \eqref{eq-q-a} and \eqref{eq-r-a} are now given by
\begin{gather*}
\text{r.h.s.~\eqref{eq-q-a}}=c\sum\mysub{B}\left(q\mysub{B'}-q\mysub{B''}\right),
\qquad
\text{r.h.s.~\eqref{eq-r-a}}=c\sum\mysub{B}\left(r\mysub{B'}-r\mysub{B''}\right)
\end{gather*}
and are identically zero because of the cyclic character of $B'$ and $B''$ (see Table~\ref{notation}).
However, substitutions \eqref{ansatz-Q} and \eqref{ansatz-R} are not so easy as it may appear.
The problem lies in the fact that the consistency of equations \eqref{ansatz-Q} and \eqref{ansatz-R} with the
lattice translations implies some restrictions on the functions $f\mysub{AB}$.
To expose them, let us apply to \eqref{ansatz-Q} with $B=U$ the shift in the `east' direction, $\tdhlT{E}$,
def\/ined by
\begin{gather*}
\tdhlT{E}f\mysub{A}=f\mysub{E},
\qquad
\tdhlT{E}f\mysub{W}=f\mysub{A},
\qquad
\tdhlT{E}f\mysub{L}=f\mysub{S},
\qquad
\text{etc}.
\end{gather*}
By simple algebra one can obtain that functions $f\mysub{AB}$ have to meet the 
condition
\begin{gather}
f\mysub{AR} \tdhlT{E}f\mysub{AU}=-c^{2},
\label{eq-ftf}
\end{gather}
which in terms of $u$ and $p$ is given by
\begin{gather}
J_{1}J_{2}\exp\left(u\mysub{A}-u\mysub{R}+u\mysub{E}-u\mysub{N}\right)=-c^{2}\sqrt{p\mysub{A}p\mysub{R}
p\mysub{E}p\mysub{N}}.
\label{eq-up-a}
\bar{}\end{gather}
Thus, we have replaced equations \eqref{eq-q-a} and \eqref{eq-r-a} with new ones, \eqref{ansatz-Q} and
\eqref{ansatz-R}, together with~\eqref{eq-up-a}.

\subsection{Rebuilding (\ref{eq-p-a})}

Ansatz \eqref{ansatz-Q} and \eqref{ansatz-R} not only enables to `solve' equation \eqref{eq-q-a} and
\eqref{eq-r-a} but also gives us possibility to simplify the remaining f\/ield equation \eqref{eq-p-a}.

With the help of \eqref{ansatz-Q} and \eqref{ansatz-R} one can present the right-hand side of \eqref{eq-p-a} as
\begin{gather*}
\text{r.h.s.~\eqref{eq-p-a}}=\mathcal{X}\mysub{A}+\mathcal{Y}\mysub{A}
\end{gather*}
with
\begin{gather*}
\mathcal{X}\mysub{A}=\frac{c}{2}\sum\mysub{B}\varepsilon\mysub{B}\left(\{q,r\}\mysub{AB'}-\{q,r\}
\mysub{AB''}\right),
\qquad
\mathcal{Y}\mysub{A}=p\mysub{A}\sum\mysub{B}f\mysub{AB}\varepsilon\mysub{B},
\end{gather*}
where we use the shorthand
\begin{gather*}
\{x,y\}\mysub{AB}=x\mysub{A}y\mysub{B}+x\mysub{B}y\mysub{A}.
\end{gather*}
The f\/irst part, $\mathcal{X}\mysub{A}$, after substituting $\varepsilon\mysub{B}$, becomes
\begin{gather*}
\mathcal{X}\mysub{A}=c\left(\{q,r\}\mysub{AE}-\{q,r\}\mysub{AW}\right)
=c\left(\tdhlT{E}-1\right)\{q,r\}\mysub{AW}.
\end{gather*}
The last expression indicates that there is a~possibility to reduce the order of our equation by presenting
the right-hand side of \eqref{eq-p-a} as $\left(\tdhlT{E}-1\right)\mathcal{S}\mysub{A}$ and solving
$\mathcal{S}\mysub{A}=\text{const}$.
To do this, we rewrite $\mathcal{Y}\mysub{A}$ with the help of \eqref{eq-ftf} as
\begin{gather*}
\mathcal{Y}\mysub{A}=p\mysub{A}\left(f\mysub{AU}-f\mysub{AL}\right)+c^{2}\;\tdhlT{E}\;p\mysub{W}
\left(f\mysub{AL}^{-1}-f\mysub{AU}^{-1}\right)
\end{gather*}
and note that we can achieve our goal by imposing the condition
\begin{gather*}
p\mysub{A}f\mysub{AU}f\mysub{AL}=-c^{2}\,p\mysub{W}.
\end{gather*}
This equation, after substituting $f\mysub{AB}$ and applying $\tdhlT{E}$, becomes
\begin{gather*}
J_{1}J_{2}\exp\left(u\mysub{S}-u\mysub{N}\right)=-c^{2}p\mysub{A}\sqrt{p\mysub{S}p\mysub{N}}
\end{gather*}
and converts \eqref{eq-p-a} into
\begin{gather}
c\{q,r\}\mysub{AW}+p\mysub{A}\left(f\mysub{AL}-f\mysub{AU}\right)=\lambda=\text{const}.
\label{eq-p-b}
\end{gather}
To summarize, at this stage the f\/ield equations can be written as the~system
\begin{gather*}
f\mysub{AL}\left(q\mysub{A}-q\mysub{L}\right)=c\left(q\mysub{W}-q\mysub{S}\right),
\qquad
f\mysub{AL}\left(r\mysub{A}-r\mysub{L}\right)=c\left(r\mysub{W}-r\mysub{S}\right),
\\
c\{q,r\}\mysub{AW}+p\mysub{A}\left(f\mysub{AL}-f\mysub{AU}\right)=\lambda
\end{gather*}
(here we write equations \eqref{ansatz-Q} and \eqref{ansatz-R} with $B=L$, keeping in mind that all the rest
can be obtained by lattice shifts) together with
\begin{gather}
J_{1}J_{2}\exp\left(u\mysub{L}-u\mysub{A}+u\mysub{S}-u\mysub{W}\right)=-c^{2}\sqrt{p\mysub{L}p\mysub{A}
p\mysub{S}p\mysub{W}},
\nonumber
\\
J_{1}J_{2}\exp\left(u\mysub{S}-u\mysub{N}\right)=-c^{2}p\mysub{A}\sqrt{p\mysub{S}p\mysub{N}}.
\label{syst-restr}
\end{gather}

\subsection{Tau-functions}

It turns out that one can solve equations \eqref{syst-restr} \emph{explicitly}, by introducing proper
\emph{parametrization} of the functions $u$ and $p$.
Omitting the technical details, we present here the results.

By easy calculation one can check that functions
\begin{gather}
u\mysub{A}=w\mysub{A}+\frac{1}{2}\ln\frac{\tau\mysub{S}}{\tau\mysub{N}},
\qquad
p\mysub{A}=\frac{\tau\mysub{N}\tau\mysub{S}}{\tau\mysub{A}^{2}},
\label{up-tau}
\end{gather}
where $w$ is a~linear function of $\vec{r}$ solve \eqref{syst-restr} identically provided that $w$
satisf\/ies
\begin{gather}
J_{1}J_{2}\exp\left(w\mysub{S}-w\mysub{N}\right)=-c^{2}.
\label{eq-up-w}
\end{gather}
The role of the function $w$, which in terms of $\vec{r}$ can be written as
$w=\left(\vec{k},\vec{r}\right)$ with a~constant vector $\vec{k}$, is that it describes (in the case of
bounded tau-functions $\tau$) the asymptotics of solutions: 
\begin{gather}
u\sim\left(\vec{k},\vec{r}\right)
\qquad
\text{as}
\qquad
|\vec{r}|\to\infty.
\label{bc-u}
\end{gather}
Substituting $u\mysub{A}$ and $p\mysub{A}$ one can rewrite $f\mysub{AB}$ as
\begin{gather}
f\mysub{AB}=J\mysub{B}\xi\mysub{B}\frac{\tau\mysub{A}\tau\mysub{B}}{\tau\mysub{B'}\tau\mysub{B''}}.
\label{def-f-tau}
\end{gather}
Here $J\mysub{B}$ and $\xi\mysub{B}$ are given in Table~\ref{notation}, the constants $\xi_{1,2}$ are
def\/ined by 
\begin{gather*}
\xi_{i}=\exp\left\{{-} (\vec{k},\vec{\delta}_{i} )\right\},
\qquad
i=1,2,
\end{gather*}
and are related by
\begin{gather*}
J_{1}J_{2} \xi_{1}\xi_{2}=-c^{2},
\end{gather*}
which ensures that \eqref{eq-up-w} is met.

The structure of equations \eqref{ansatz-Q}, \eqref{ansatz-R} and \eqref{def-f-tau} suggests the
representation of $q\mysub{A}$ and $r\mysub{A}$,
\begin{gather*}
q\mysub{A}=\frac{\sigma\mysub{A}}{\tau\mysub{A}},
\qquad
r\mysub{A}=\frac{\rho\mysub{A}}{\tau\mysub{A}},
\end{gather*}
where real tau-functions $\tau$ and complex ones, $\sigma$ and $\rho$, are related by
\begin{gather*}
\tau\mysub{A}^{2}=\sigma\mysub{A}\rho\mysub{A}+\tau\mysub{N}\tau\mysub{S}
\end{gather*}
which is the consequence of the identity $p\mysub{A}=1-q\mysub{A}r\mysub{A}$.
This representation immediately converts~\eqref{ansatz-Q} and~\eqref{ansatz-R} into bilinear equations
\begin{gather*}
0=[\sigma,\tau]\mysub{AB}-\mu\mysub{B}[\sigma,\tau]\mysub{B'B''},
\qquad
0=[\rho,\tau]\mysub{AB}-\mu\mysub{B}[\rho,\tau]\mysub{B'B''},
\end{gather*}
where we utilize another shorthand,
\begin{gather*}
[x,y]\mysub{AB}=x\mysub{A}y\mysub{B}-x\mysub{B}y\mysub{A},
\end{gather*}
and use, instead of $J\mysub{B}$ and $\xi\mysub{B}$, constants $\mu\mysub{B}$ collected in
Table~\ref{notation}, with
\begin{gather}
\mu=\frac{c}{J_{1}\xi_{1}}=-\frac{J_{2}\xi_{2}}{c}.
\label{eq-mu}
\end{gather}
Finally, one can see that the right-hand side of the last f\/ield equation, \eqref{eq-p-b}, becomes
\begin{gather*}
\text{r.h.s.~\eqref{eq-p-b}}=\frac{c}{\tau\mysub{A}\tau\mysub{W}}\left(\{\sigma,\rho\}\mysub{AW}
+\mu\tau\mysub{U}\tau\mysub{S}+\mu^{-1}\tau\mysub{N}\tau\mysub{L}\right)
\end{gather*}
that leads to the bilinearization of \eqref{eq-p-b}:
\begin{gather*}
\{\sigma,\rho\}\mysub{AW}+\mu\tau\mysub{U}\tau\mysub{S}+\mu^{-1}\tau\mysub{L}\tau\mysub{N}=\frac{\lambda}{c}
 \tau\mysub{A}\tau\mysub{W}.
\end{gather*}
In what follows we restrict ourselves to the case of bounded $u-w$, which corresponds to
\begin{gather*}
\lambda=c\left(\mu+\mu^{-1}\right).
\end{gather*}
Thus, we have proved the following
\begin{prop}
\label{prop-bilin}
A wide range of solutions for the field equations \eqref{euler-u} and \eqref{euler-phi} can be obtained
from the  bilinear  system
\begin{gather}
0=\tau\mysub{A}^{2}-\tau\mysub{N}\tau\mysub{S}-\sigma\mysub{A}\rho\mysub{A},
\nonumber
\\
0=[\sigma,\tau]\mysub{AL}+\mu[\sigma,\tau]\mysub{SW},
\nonumber
\\
0=[\rho,\tau]\mysub{AL}+\mu[\rho,\tau]\mysub{SW},
\nonumber
\\
0=\{\sigma,\rho\}\mysub{AW}+\mu\tau\mysub{U}\tau\mysub{S}+\mu^{-1}\tau\mysub{L}\tau\mysub{N}
-\left(\mu+\mu^{-1}\right)\tau\mysub{A}\tau\mysub{W}.
\label{syst-bilin}
\end{gather}
\end{prop}
This system is the one we were looking for: a~bilinear system providing solutions for the f\/ield equations
\eqref{euler-u} and \eqref{euler-phi}.
However, these equations, except the f\/irst one, are four- and f\/ive-site ones, which makes their
solution rather cumbersome.
It is possible to prove directly that solutions presented below (see section~\ref{sec-sol}) satisfy
\eqref{syst-bilin}.
However we take another way and \emph{split}, in the next section, this system in a~set of standard
Hirota-like three-term equations.
Of course, such splitting narrows the class of solutions.
Nevertheless, this class is rather rich and contains the soliton solutions that we want to obtain.

\section[Reduction of (\ref{syst-bilin}) to the Ablowitz-Ladik equations]{Reduction of (\ref{syst-bilin}) to the Ablowitz--Ladik equations}
\label{sec-alh}

To explain the idea behind the splitting we are going to do, it seems reasonable to pass from the vertex
notation to one based on the shifts/translations $\tdhlT{}$,
\begin{gather*}
f\mysub{A}=f,
\qquad
f\mysub{W}=\tdhlshifted{W}{f},
\qquad
f\mysub{S}=\tdhlshifted{S}{f},
\qquad
\text{etc}.
\end{gather*}
In this notation, the bilinear system \eqref{syst-bilin} can be rewritten as
\begin{gather}
0=\tau^{2}-\rho\sigma-\tdhlShifted{S}{\tau}\tdhlShifted{N}{\tau},
\nonumber
\\
0=\tau\tdhlShifted{L}{\sigma}-\sigma\tdhlShifted{L}{\tau}+\mu\left[\tdhlShifted{S}{\tau}
\tdhlShifted{W}{\sigma}-\tdhlShifted{S}{\sigma}\tdhlShifted{W}{\tau}\right],
\nonumber
\\
0=\tau\tdhlShifted{L}{\rho}-\rho\tdhlShifted{L}{\tau}+\mu\left[\tdhlShifted{S}{\tau}\tdhlShifted{W}
{\rho}-\tdhlShifted{S}{\rho}\tdhlShifted{W}{\tau}\right],
\nonumber
\\
0=\sigma\tdhlShifted{W}{\rho}+\rho\tdhlShifted{W}{\sigma}-\left(\mu+\mu^{-1}
\right)\tau\tdhlShifted{W}{\tau}+\mu\tdhlShifted{S}{\tau}\tdhlShifted{U}{\tau}+\mu^{-1}\tdhlShifted{N}{\tau}
\tdhlShifted{L}{\tau}.
\label{eqs-to-solve}
\end{gather}
Each point of both direct and dual lattices can be reached by a~composition of \emph{two} basic shifts, say,
$\tdhlshifted{S}{}$ and $\tdhlshifted{W}{}$,
\begin{gather*}
\tdhlT{N}=\tdhlT{S}^{-1},
\qquad
\tdhlT{L}=\tdhlT{S}\tdhlT{W},
\qquad
\tdhlT{U}=\tdhlT{S}^{-1}\tdhlT{W},
\qquad
\text{etc}.
\end{gather*}
and hence all equations can be presented in terms of these two translations only.
The trick that leads to the reduction to three-term equations is to use \emph{three} shifts as a~basic
system.
One of them is, say, $\tdhlT{S}$ while two more shifts come from the \emph{splitting} of $\tdhlT{L}$,
\begin{gather*}
\tdhlT{L}=\tdhlT{X}\tdhlT{Y},
\end{gather*}
or, alternatively, $\tdhlT{S}\tdhlT{W}=\tdhlT{X}\tdhlT{Y}$.
Application of this construction to our equations leads to
\begin{prop}
\label{prop-alh}
A wide range of solutions for the field equations \eqref{euler-u} and \eqref{euler-phi} can be obtained
from the bilinear system of Hirota-like equations
\begin{gather}
0=\tau^{2}-\rho\sigma-\tdhlShifted{S}{\tau}\tdhlShifted{N}{\tau},
\label{alh-a}
\\
0=\tau\tdhlShifted{X}{\tau}-\sigma\tdhlShifted{X}{\rho}-\tdhlShifted{S}{\tau}\tdhlShifted{NX}{\tau},
\label{alh-b}
\\
0=\mu\tdhlShifted{S}{\sigma}\tdhlShifted{X}{\rho}-\tdhlShifted{S}{\tau}\tdhlShifted{X}{\tau}
+\tau\tdhlShifted{SX}{\tau},
\label{alh-c}
\\
0=\tau\tdhlShifted{Y}{\tau}-\rho\tdhlShifted{Y}{\sigma}-\tdhlShifted{S}{\tau}\tdhlShifted{NY}{\tau},
\label{alh-d}
\\
0=\mu\tdhlShifted{S}{\rho}\tdhlShifted{Y}{\sigma}-\tdhlShifted{S}{\tau}\tdhlShifted{Y}{\tau}
+\tau\tdhlShifted{SY}{\tau}.
\label{alh-e}
\end{gather}
\end{prop}

One can f\/ind a~proof of this statement in Appendix~\ref{app-proof}.
Geometrically, this constructions can be interpreted as if we considered our plane as a~part of
three-dimensional lattice and presented our bilinear equations as a~projection of more simple
three-dimensional system.

Equations \eqref{alh-a}--\eqref{alh-e} are closely related to the Ablowitz--Ladik system~\cite{AL75,AL76}.
Indeed, if we think of the tau-functions as depending on a~discrete index, $n$, and two inf\/inite sets of
`times', $\left\{t_{j}\right\}_{j=1}^{\infty}$ and $\left\{\bar{t}_{k}\right\}_{k=1}^{\infty}$, and
identify $\tdhlT{S,N}$ with the shifts $n\to n\pm1$ and $\tdhlT{X,Y}$ with the Miwa shifts $t_{j}\to
t_{j}\pm i\xi^{j}/j$ and $\bar{t}_{k}\to\bar{t}_{k}\pm i\eta^{k}/k$, then \eqref{alh-a}--\eqref{alh-e},
considered as functional equations, describe the positive and negative f\/lows of the ALH (see~\cite{V02}).
Thus, the calculations of the previous section could be replaced with the statement that \emph{each}
solution for the ALH provides a~solution for the 2DTHL (not only the soliton ones that we derive below).
However, it seems that such approach is not the best in the case of the model we study.
In principle, it is possible to rewrite equations \eqref{alh-a}--\eqref{alh-e} in terms of $u_{\vec{r}}$ and
$\vec{\phi}_{\vec{r}}$, but the resulting equations (which we do not write here) are rather cumbersome and
can hardly give clear understanding of, say, the origin of $u$-$p$ relations \eqref{syst-restr}.

\section{Soliton solutions for (\ref{alh-a})--(\ref{alh-e})}
\label{sec-sol}

In this section we derive the soliton solutions for the bilinear system \eqref{alh-a}--\eqref{alh-e}.
This will be done in two steps.
First we solve it without taking into account the condition $r=-q^{*}$ and, secondly, f\/ind the
restrictions that should be imposed on the parameters of the solutions to meet it.

Since equations \eqref{alh-a}--\eqref{alh-e} are nothing but the Ablowitz--Ladik system, we use some of the
results of papers~\cite{AL75,AL76}, namely the structure of the soliton solutions without developing the
inverse scattering transform from scratch.

The main blocks for constructing the $N$-soliton solutions are the matrices that satisfy the `almost
rank-one' conditions
\begin{gather}
\tdhlL\tdhlmatrix{A}-\tdhlmatrix{A}\tdhlR=\tdhlket{1}\tdhlbra{a},
\qquad
\tdhlR\tdhlmatrix{B}-\tdhlmatrix{B}\tdhlL=\tdhlket{1}\tdhlbra{b}.
\label{rank-one}
\end{gather}
Here $\tdhlmatrix{L}$ and $\tdhlmatrix{R}$ are constant diagonal matrices, $\tdhlket{1}$ is a~constant
$N$-component column, $\tdhlbra{a}$~and~$\tdhlbra{b}$ are $N$-component rows depending on the coordinates
of the problem.

We look for solutions of the form, similar to the form of soliton solutions for the Ablowitz--Ladik model:
\begin{gather}
\tau=\det\left|\tdhlmatrix{1}-\tdhlmatrix{A}\tdhlmatrix{B}\right|=\det\left|\tdhlmatrix{1}-\tdhlmatrix{B}
\tdhlmatrix{A}\right|
\label{sols-tau}
\end{gather}
and
\begin{gather}
q=\tdhlbra{a}\tdhlmatrix{F}\tdhlR^{-1}\tdhlket{1},
\qquad
r=-\tdhlbra{b}\tdhlL^{-1}\tdhlmatrix{G}\tdhlket{1}.
\label{sols-qr}
\end{gather}
Here $\tdhlmatrix{1}$ is the $N\times N$ unit matrix,
\begin{gather*}
\tdhlmatrix{F}=\left(\tdhlmatrix{1}-\tdhlmatrix{B}\tdhlmatrix{A}\right)^{-1},
\qquad
\tdhlmatrix{G}=\left(\tdhlmatrix{1}-\tdhlmatrix{A}\tdhlmatrix{B}\right)^{-1}
\end{gather*}
and $\tdhlbra{a}\tdhlmatrix{M}\tdhlket{1}$ is the standard row-matrix-column product:
$\tdhlbra{a}\tdhlmatrix{M}\tdhlket{b}=\sum\limits_{jk}a_{j}M_{jk}b_{k}$,
where $a_{j}$, $M_{jk}$ and $b_{k}$ are components of a~row $\tdhlbra{a}$, a~matrix $\tdhlmatrix{M}$ and
a~column $\tdhlket{b}$.

The second part of the `solitonic ansatz' is that the action of the shifts $\tdhlT{S,X,Y,\dots}$ can be
implemented as the right multiplication by diagonal constant matrices. 
The structure of these matrices, that are rational functions of $\tdhlL$ and $\tdhlR$, can be obtained,
again, from~\cite{AL75,AL76}.
The resulting formulae can be written as follows:
\begin{prop}
Soliton solutions for \eqref{alh-a}--\eqref{alh-e} are given by
\begin{gather*}
\tau=\det\left|\tdhlmatrix{1}-\tdhlmatrix{A}\tdhlmatrix{B}\right|=\det\left|\tdhlmatrix{1}-\tdhlmatrix{B}
\tdhlmatrix{A}\right|
\end{gather*}
and
\begin{gather*}
\sigma=\tau\tdhlbra{a}\tdhlmatrix{F}\tdhlR^{-1}\tdhlket{1},
\qquad
\rho=-\tau\tdhlbra{b}\tdhlL^{-1}\tdhlmatrix{G}\tdhlket{1}
\end{gather*}
with the following actions of the $\tdhlT{}$-shifts:
\begin{alignat}{3}
& \tdhlshifted{S}{\tdhlmatrix{A}}=\tdhlmatrix{A}\tdhlR,
\qquad && \tdhlshifted{S}{\tdhlmatrix{B}}=\tdhlmatrix{B}\tdhlL^{-1}, &\nonumber\\
& \tdhlshifted{X}{\tdhlmatrix{A}}=\tdhlmatrix{A}\left(\tdhlmatrix{1}+\mu\tdhlR\right),
\qquad && \tdhlshifted{X}{\tdhlmatrix{B}}=\tdhlmatrix{B}\left(\tdhlmatrix{1}+\mu\tdhlL\right)^{-1}, & \label{sols-shift-SXY}\\
& \tdhlshifted{Y}{\tdhlmatrix{A}}=\tdhlmatrix{A}\left(\tdhlmatrix{1}+\mu\tdhlR^{-1}\right)^{-1},
\qquad &&
 \tdhlshifted{Y}{\tdhlmatrix{B}}=\tdhlmatrix{B}\left(\tdhlmatrix{1}+\mu\tdhlL^{-1}\right) &\nonumber
\end{alignat}
and
\begin{alignat}{3}
& \tdhlshifted{S}{\tdhlbra{a}}=\tdhlbra{a}\tdhlR,
\qquad && \tdhlshifted{S}{\tdhlbra{b}}=\tdhlbra{b}\tdhlL^{-1},&\nonumber\\
& \tdhlshifted{X}{\tdhlbra{a}}=\tdhlbra{a}\left(\tdhlmatrix{1}+\mu\tdhlR\right),
\qquad && \tdhlshifted{X}{\tdhlbra{b}}=\tdhlbra{b}\left(\tdhlmatrix{1}+\mu\tdhlL\right)^{-1}, & \label{sols-shifts-bra}\\
& \tdhlshifted{Y}{\tdhlbra{a}}=\tdhlbra{a}\left(\tdhlmatrix{1}+\mu\tdhlR^{-1}\right)^{-1},
\qquad && \tdhlshifted{Y}{\tdhlbra{b}}=\tdhlbra{b}\left(\tdhlmatrix{1}+\mu\tdhlL^{-1}\right).& \nonumber
\end{alignat}
\end{prop}

This proposition is proved in Appendix~\ref{app-ver}.

After having derived the `general' soliton solutions for \eqref{alh-a}--\eqref{alh-e}, we have to ensure the
`physical' involution $r=-q^{*}$, or
\begin{gather}
\tau^{*}=\tau,
\qquad
\rho=-\sigma^{*}
\label{tau-convolution}
\end{gather}
bearing in mind that
\begin{gather*}
\mu^{*}=\epsilon\mu,
\qquad
\epsilon=\pm1,
\end{gather*}
which follows from the fact that $\mu^{2}=-J_{2}\xi_{2}/J_{1}\xi_{1}$ is real, but possibly negative,
number (see~\eqref{eq-mu}).

Omitting rather straightforward calculations, we present here the following results: the relationships
between the matrices $\tdhlmatrix{A}$ and $\tdhlmatrix{B}$, and hence between the rows~$\tdhlbra{a}$ and~$\tdhlbra{b}$, that ensure~\eqref{tau-convolution} are
\begin{gather*}
\tdhlL^{*}=\epsilon\tdhlR^{-1},
\qquad
\tdhlR^{*}=\epsilon\tdhlL^{-1}
\end{gather*}
together with
\begin{gather}
\tdhlmatrix{A}^{*}=-\tdhlR\tdhlmatrix{B}\tdhlL^{-1},
\qquad
\tdhlmatrix{B}^{*}=-\tdhlL\tdhlmatrix{A}\tdhlR^{-1}
\label{inv-AB}
\end{gather}
and
\begin{gather}
\tdhlbra{a^{*}}=\epsilon\tdhlbra{b}\tdhlL^{-2}.
\label{inv-ab}
\end{gather}
Restrictions \eqref{inv-AB}, \eqref{inv-ab} can be resolved by introducing, instead of $\tdhlmatrix{A}$ and
$\tdhlmatrix{B}$, one matrix $\tdhlmatrix{C}$,
\begin{gather}
\tdhlmatrix{A}=\tdhlmatrix{C}\tdhlR,
\qquad
\tdhlmatrix{B}=-\epsilon\tdhlR^{-1}\tdhlmatrix{C}^{*},
\label{inv-def-C}
\end{gather}
and 
the single row $\tdhlbra{c}$ instead of $\tdhlbra{a}$ and $\tdhlbra{b}$,
\begin{gather}
\tdhlbra{a}=\tdhlbra{c}\tdhlR,
\qquad
\tdhlbra{b}=\tdhlbra{c^{*}}\tdhlL.
\label{inv-def-c}
\end{gather}
Def\/initions \eqref{inv-def-C} and \eqref{inv-def-c} enable to present $\tdhlmatrix{F}$ and $\tdhlmatrix{G}$
as
\begin{gather*}
\tdhlmatrix{F}=\tdhlR^{-1}\tdhlmatrix{H}\tdhlR,
\qquad
\tdhlmatrix{G}=\tdhlmatrix{H}^{*},
\end{gather*}
where
\begin{gather*}
\tdhlmatrix{H}=\left(\tdhlmatrix{1}+\epsilon\tdhlmatrix{C}^{*}\tdhlmatrix{C}\right)^{-1}
\end{gather*}
and to rewrite \eqref{sols-tau} and \eqref{sols-qr} as
\begin{gather*}
\tau=\det\left|\tdhlmatrix{1}+\epsilon\tdhlmatrix{C}^{*}\tdhlmatrix{C}\right|
\end{gather*}
and
\begin{gather*}
q=\tdhlbra{c}\tdhlmatrix{H}\tdhlket{1},
\qquad
r=-\tdhlbra{c^{*}}\tdhlmatrix{H}^{*}\tdhlket{1},
\end{gather*}
which clearly demonstrates the fulf\/ilment of \eqref{tau-convolution}.

\section{Solitons of the 2DTHL}

Now we have all necessary to present the $N$-soliton solutions for equations \eqref{euler-u},
\eqref{euler-phi}, i.e.\ to write down the expressions describing the solitons of the 2DTHL.

First we have to return from the next-neighbour notation to the `absolute' one noting that for any lattice
vector,
\begin{gather*}
\vec{r}=m_{1}\vec{\delta}_{1}+m_{2}\vec{\delta}_{2},
\end{gather*}
and any function $f_{\vec{r}}$
\begin{gather*}
f_{\vec{r}}=\tdhlT{R}^{m_{1}}\tdhlT{U}^{m_{2}}f,
\end{gather*}
where $f$ is the value of the function at some f\/ixed point.
Recalling that $\tdhlT{L}=\tdhlT{X}\tdhlT{Y}$ and that
\begin{gather*}
\tdhlT{R}=\tdhlT{L}^{-1},
\qquad
\tdhlT{U}=\tdhlT{L}\tdhlT{S}^{-2}
\end{gather*}
one can introduce the matrices $\tdhlMatrix{C}{\vec{r}}$ and the rows $\tdhlBra{c}{\vec{r}}$ by
\begin{gather*}
\tdhlMatrix{C}{\vec{r}}=\tdhlMatrix{C}{}\,\tdhlMatrix{M}{1}^{m_{1}}\tdhlMatrix{M}{2}^{m_{2}},
\qquad
\tdhlBra{c}{\vec{r}}=\tdhlBra{c}{}\,\tdhlMatrix{M}{1}^{m_{1}}\tdhlMatrix{M}{2}^{m_{2}},
\end{gather*}
where, as follows from \eqref{sols-shift-SXY},
\begin{gather*}
\tdhlMatrix{M}{1}=\left(\tdhlmatrix{1}+\mu\tdhlR^{-1}\right)\left(\tdhlmatrix{1}+\mu\tdhlR\right)^{-1},
\qquad
\tdhlMatrix{M}{2}=\left(\tdhlR^{-1}+\mu\tdhlmatrix{1}\right)\left(\tdhlR+\mu\tdhlmatrix{1}\right)^{-1}
\end{gather*}
with \textit{constant} $\tdhlMatrix{C}{}$ and $\tdhlBra{c}{}$ related by
\begin{gather*}
\tdhlL\tdhlmatrix{C}-\tdhlmatrix{C}\tdhlR=\tdhlket{1}\tdhlbra{c}
\qquad \text{or}\qquad
\left(\tdhlMatrix{C}{}\right)_{jk}=\frac{c_{k}}{L_{j}-R_{k}}=\frac{\epsilon R_{j}^{*}c_{k}}{1-\epsilon R_{j}
^{*}R_{k}}.
\end{gather*}
The parameter $\mu$ depends on vector $\vec{k}$ that def\/ines the asymptotics of solutions (see
\eqref{bc-u}),
\begin{gather*}
\mu=\mu(\vec{k})=\sqrt{-\frac{J_{2}}{J_{1}}}\exp\left[\tfrac 12 \left(\vec{k},\vec{\delta}_{1}-\vec{\delta}_{2}
\right)\right]
\end{gather*}
while the constant $\epsilon$ is given by
\begin{gather*}
\epsilon=-\mathop{\text{sign}}J_{1}J_{2}.
\end{gather*}
In this notation, the main formulae of the previous section become
\begin{gather}
q_{\vec{r}}=\tdhlBra{c}{\vec{r}}\tdhlMatrix{H}{\vec{r}}\tdhlket{1},
\label{final-q}
\end{gather}
where
\begin{gather*}
\tdhlMatrix{H}{\vec{r}}=\left(\tdhlmatrix{1}+\epsilon\tdhlMatrix{C}{\vec{r}}^{*}\tdhlMatrix{C}{\vec{r}}
\right)^{-1}
\qquad \text{and}\qquad
\tau_{\vec{r}}=\det\left|\tdhlmatrix{1}+\epsilon\tdhlMatrix{C}{\vec{r}}^{*}\tdhlMatrix{C}{\vec{r}}\right|.
\end{gather*}
Finally, one can present the soliton solutions of the model considered in this paper as follows: the
vectors $\phi_{\vec{r}}$ are given by
\begin{gather*}
\vec{\phi}_{\vec{r}}=\left(
\begin{matrix}\sin\theta_{\vec{r}}\cos\varphi_{\vec{r}}
\\
\sin\theta_{\vec{r}}\sin\varphi_{\vec{r}}
\\
\cos\theta_{\vec{r}}
\end{matrix}
\right)
\end{gather*}
with
\begin{gather*}
\theta_{\vec{r}}=\arcsin\frac{\left|q_{\vec{r}}\right|}{1+\left|q_{\vec{r}}\right|^{2}},
\qquad
\varphi_{\vec{r}}=\arg q_{\vec{r}},
\end{gather*}
where $q_{\vec{r}}$ are def\/ined by~\eqref{final-q} while
\begin{gather*}
u_{\vec{r}}=\left(\vec{k},\vec{r}\right)+\frac{1}{2}\ln\frac{\tau_{\vec{r}}^{+}}{\tau_{\vec{r}}^{-}}
\end{gather*}
with
\begin{gather*}
\tau_{\vec{r}}^{\pm}=\tdhlT{S}^{\pm1}\tau_{\vec{r}}=\det\left|\tdhlmatrix{1}+\epsilon\tdhlMatrix{C}{\vec{r}}
^{*}\tdhlL^{\mp1}\tdhlMatrix{C}{\vec{r}}\tdhlR^{\pm1}\right|.
\end{gather*}

\section{Conclusion}

In this paper, we have presented the nonlinear 2D lattice and have obtained its $N$-soliton solutions.
To conclude, we would like to give some comments related to the proposed model, the method we used to solve
it and to outline the possible continuation of this work.

The key moment in the bilinearization of the f\/ield equations were equations \eqref{ansatz-Q} and
\eqref{ansatz-R}.
Since we did not expect to obtain the general solution and the objective was do derive some particular
ones, we have not studied it in details.
We used it as an \emph{ansatz}, a~trick that helps us to achieve our goal (even despite the loss of the
generality).
However this substitution, nonlocal and rather cumbersome when rewritten in terms of the original
variables, $u_{\vec{r}}$ and $\vec{\phi}_{\vec{r}}$, surely needs to be studied in a~more detailed way.
In some sense, equations \eqref{ansatz-Q} and \eqref{ansatz-R} can be viewed as an quadrilateral version of
the star-triangle transformation (see~\cite{B82} and references therein).
To our opinion, it may prove useful in the studies of other two-dimensional lattice models.

Considering the loss of generality, we have to admit that our approach, is surely a~reduction.
First, the $u$-$\vec{\phi}$ `mixing' that we made by introducing the tau-functions by \eqref{vec-scal},
\eqref{def-p} and \eqref{up-tau} narrows the class of solutions that we can obtain.
For example, the `frozen spin' conf\/igurations $\vec{\phi}_{\vec{r}}=(0,0,1)^{T}$ are described by the
Hirota-like model \eqref{energy-Toda} whose solutions hardly can be obtained in the framework of the method
of this paper because the restriction $\vec{\phi}\mysub{A}=(0,0,1)^{T}$ implies $q\mysub{A}=0$ and
$p\mysub{A}=1$ which drastically simplif\/ies all equations from Propositions~\ref{prop-bilin}
and~\ref{prop-alh} leavig us with almost trivial solutions for the model \eqref{energy-Toda}.
A similar ef\/fect occurs when we split the second-order equations into the f\/irst-order system (see
section~\ref{sec-alh}), which can be viewed as the second step of the reduction.
To illustrate this fact we would like to note, for example, that model \eqref{energy-vec} considered in this
paper admits a~non-trivial one-dimensional reduction $u_{\vec{r}}=u_{\vec{r}+\vec{\delta}_{1}}$,
$\vec{\phi}_{\vec{r}}=\vec{\phi}_{\vec{r}+\vec{\delta}_{1}}$ whose most interesting solutions, again,
cannot obtained directly from the ones derived above: one can easily see from equations
\eqref{alh-a}--\eqref{alh-e} that the reduction $\tdhlshifted{L}{u}=u$,
$\tdhlshifted{L}{\vec{\phi}}=\vec{\phi}$ leads to $\tdhlshifted{S}{\tau}=\text{const}\cdot\tau$ which
means that equations from Proposition~\ref{prop-alh} provide only almost trivial solutions for the
one-dimensional problem.
To summarize, not all solutions for~\eqref{euler-u} and~\eqref{euler-phi}, can be obtained from
\eqref{alh-a}--\eqref{alh-e}.
However the class of solutions described by \eqref{alh-a}--\eqref{alh-e} is rather rich and seems to include,
not strictly speaking, almost all essentially two-dimensional solutions that can be written explicitly.

As it was said above, equations \eqref{alh-a}--\eqref{alh-e} belong to the ALH.
The fact that ALH-equations lead to various Toda-like ones is not new.
For example, it has been shown in~\cite{V95} that equations describing the famous two-dimensional Toda
lattice can be splitted into ones for the simplest ALH f\/lows.
Second example is the recent paper~\cite{V13}, where the author discusses the relationships between the ALH
and the relativistic Toda~\cite{BR89a,R90} and the two-dimensional Volterra~\cite{LSS80} models.
On the other hand, there is a~number of papers that demonstrate the links between the ALH and
Heisenberg-like models.
Probably the f\/irst such example is Ishimori classical spin chain~\cite{Y82} which is gauge equivalent to
the discrete nonlinear Schr\"odinger equation (one of the most well-studied equations of the ALH).
Another example one can f\/ind in~\cite{PV11}, where the authors study a~model of Landau--Lifshitz f\/ields
interacting in a~Heisenberg-like way.
However there is an important distinction between the present and the above-cited works.
The case is that all models from~\cite{BR89a,Y82, LSS80,PV11,R90,V13, V95} are \emph{chains}.
Indeed, even the two-dimensional Toda and Volterra models are two-dimensional in the sense that they are
equations for the functions of \emph{two continuous} but only \emph{one discrete} variables, while the
2DTHL is two-dimensional as a~lattice.
Thus, this paper is an attempt to extend the `reducing to the ALH' approach to a~family of 2D models whose
most elegant examples are Hirota models~\cite{H77a,H77b} (though they are usually considered in the
framework of discrete-time evolution and not as the 2D lattices).

The fact that the 2DTHL possesses the $N$-soliton solutions is a~strong evidence of the integrability of
the problem~\cite{Hiet87a,Hiet87b,Hiet87c,Hiet88, H04,NY86} (see also~\cite{HZ12,HZ11} for its application
to discrete systems).
However, the situations with the 2DTHL can be more complicated: there is a~possibility that equations from
Propositions~\ref{prop-bilin} and~\ref{prop-alh} belong to the family of the so-called `conditionally
integrable' systems introduced by Dorizzi et al.~\cite{DGRW86}.
In any case, the questions related to the integrability of the 2DTHL, its zero-curvature representation and
conserved quantities surely deserve further studies.

Another feasible continuation of this work is related to prospective applications to the theory of magnetic
systems.
Traditionally, the soliton solutions are most representative ones for any integrable system.
However, from the viewpoint of the theory of magnetism, it would be interesting to f\/ind other families of
solutions, except the \textit{non-topological} solitons presented in this paper, that correspond to more
typical magnetic structures as, e.g., domain walls and f\/inite magnetic domains.

\appendix \section{Proof of Proposition~\ref{prop-alh}}
\label{app-proof}

The proof of Proposition~\ref{prop-alh} is straightforward and consists in presenting the right-hand sides
of the equations from Proposition~\ref{prop-bilin} as combinations of the right-hand sides of the equations
from Proposition~\ref{prop-alh}.
To this end, consider the following def\/initions:
\begin{gather}
\tdhlz{0}=\tau^{2}-\rho\sigma-\tdhlShifted{S}{\tau}\tdhlShifted{N}{\tau},
\nonumber
\\
\tdhlz{1}=\tau\tdhlShifted{X}{\tau}-\sigma\tdhlShifted{X}{\rho}-\tdhlShifted{S}{\tau}\tdhlShifted{NX}
{\tau},
\label{def-z-b}
\\
\tdhlz{2}=\mu\tdhlShifted{S}{\sigma}\tdhlShifted{X}{\rho}-\tdhlShifted{S}{\tau}\tdhlShifted{X}{\tau}
+\tau\tdhlShifted{SX}{\tau},
\nonumber
\\
\tdhlz{3}=\tau\tdhlShifted{Y}{\tau}-\rho\tdhlShifted{Y}{\sigma}-\tdhlShifted{S}{\tau}\tdhlShifted{NY}
{\tau},
\nonumber
\\
\tdhlz{4}=\mu\tdhlShifted{S}{\rho}\tdhlShifted{Y}{\sigma}-\tdhlShifted{S}{\tau}\tdhlShifted{Y}{\tau}
+\tau\tdhlShifted{SY}{\tau}
\label{def-z-e}
\end{gather}
(which are shortcuts for the right-hand sides of \eqref{alh-a}--\eqref{alh-e}).
Our aim is to prove that vanishing of $\tdhlz{0}$, $\tdhlz{5}$ and $\tdhlz{6}$ implies vanishing of
\begin{gather*}
\tdhlZ{1}=\tau\tdhlShifted{L}{\sigma}-\sigma\tdhlShifted{L}{\tau}+\mu\tdhlShifted{S}{\tau}\tdhlShifted{W}
{\sigma}-\mu\tdhlShifted{S}{\sigma}\tdhlShifted{W}{\tau},
\\ 
\tdhlZ{2}=\tau\tdhlShifted{L}{\rho}-\rho\tdhlShifted{L}{\tau}+\mu\tdhlShifted{S}{\tau}\tdhlShifted{W}
{\rho}-\mu\tdhlShifted{S}{\rho}\tdhlShifted{W}{\tau},
\\ 
\tdhlZ{3}=\sigma\tdhlShifted{W}{\rho}+\rho\tdhlShifted{W}{\sigma}-\left(\mu+\mu^{-1}\right)\tau\tdhlShifted{W}{\tau}
+\mu\tdhlShifted{S}{\tau}\tdhlShifted{U}{\tau}+\mu^{-1}\tdhlShifted{N}{\tau}\tdhlShifted{L}{\tau}
\end{gather*}
(the right-hand sides of equations \eqref{eqs-to-solve}).
The f\/irst part of the proof is simple: it can be shown that $\tdhlZ{1}$ and $\tdhlZ{2}$ are linear
combinations of $\tdhlz{5}$ and $\tdhlz{6}$.
Indeed, one can check that
\begin{gather*}
\tdhlShifted{X}{\rho}\tdhlZ{1}=\tdhlShifted{L}{\tau}\tdhlz{1}-\tdhlShifted{W}{\tau}\tdhlz{2}
-\tau
\big({\mathbb T}_{\scriptscriptstyle X} {\mathfrak c}^{(1)}\big)
+
\tdhlShifted{S}{\tau}
\big({\mathbb T}_{\scriptscriptstyle NX} {\mathfrak c}^{(2)}\big),
\\
\tdhlShifted{Y}{\sigma}\tdhlZ{2}=-\tau
\big({\mathbb T}_{\scriptscriptstyle Y} {\mathfrak b}^{(1)}\big)
+\tdhlShifted{S}{\tau}
\big({\mathbb T}_{\scriptscriptstyle NY} {\mathfrak b}^{(2)}\big)
+\tdhlShifted{L}{\tau}\tdhlz{3}-\tdhlShifted{W}{\tau}\tdhlz{4}.
\end{gather*}
Thus
\begin{gather*}
\tdhlz{5}=\tdhlz{6}=0
\qquad
\Rightarrow
\qquad
\tdhlZ{1}=\tdhlZ{2}=0.
\end{gather*}
As to $\tdhlZ{3}$, the calculations are slightly more complicated and can be performed in two steps.
First, writing the system composed of \eqref{def-z-b} and shifted \eqref{def-z-e},
\begin{gather*}
\tdhlz{1}=\tau\tdhlShifted{X}{\tau}-\sigma\tdhlShifted{X}{\rho}-\tdhlShifted{S}{\tau}\tdhlShifted{NX}{\tau},
\\ 
\tdhlshifted{NX}{\tdhlz{4}}=\mu\tdhlShifted{W}{\sigma}\tdhlShifted{X}{\rho}-\tdhlShifted{W}{\tau}
\tdhlShifted{X}{\tau}+\tdhlShifted{L}{\tau}\tdhlShifted{NX}{\tau},
\end{gather*}
and eliminating $\tdhlShifted{X}{\rho}$ one arrives at
\begin{gather*}
\sigma
\big({\mathbb T}_{\scriptscriptstyle NX} {\mathfrak c}^{(2)}\big)
+\mu\tdhlShifted{W}{\sigma}\tdhlz{1}=A\tdhlShifted{X}{\tau}
-B\tdhlShifted{NX}{\tau},
\end{gather*}
where
\begin{gather*}
A=\mu\tau\tdhlShifted{W}{\sigma}-\sigma\tdhlShifted{W}{\tau},
\qquad
B=\mu\tdhlShifted{S}{\tau}\tdhlShifted{W}{\sigma}-\sigma\tdhlShifted{L}{\tau}.
\end{gather*}
In a~similar way, elimination of $\tdhlshifted{NX}{\rho}$ from
\begin{gather*}
\tdhlshifted{N}{\tdhlz{2}}=\mu\sigma\tdhlShifted{NX}{\rho}-\tau\tdhlShifted{NX}{\tau}+\tdhlShifted{N}{\tau}\tdhlShifted{X}{\tau},
\\ 
\tdhlshifted{NX}{\tdhlz{3}}=\tdhlShifted{W}{\tau}\tdhlShifted{NX}{\tau}-\tdhlShifted{W}{\sigma}
\tdhlShifted{NX}{\rho}-\tdhlShifted{U}{\tau}\tdhlShifted{X}{\tau}
\end{gather*}
leads to
\begin{gather*}
\tdhlShifted{W}{\sigma}
\big({\mathbb T}_{\scriptscriptstyle N} {\mathfrak b}^{(2)}\big) 
+\mu\sigma
\big({\mathbb T}_{\scriptscriptstyle NX} {\mathfrak c}^{(1)}\big)
=-C\tdhlShifted{X}
{\tau}+D\tdhlShifted{NX}{\tau}
\end{gather*}
with
\begin{gather*}
C=\mu\sigma\tdhlShifted{U}{\tau}-\tdhlShifted{N}{\tau}\tdhlShifted{W}{\sigma},
\qquad
D=\mu\sigma\tdhlShifted{W}{\tau}-\tau\tdhlShifted{W}{\sigma}
\end{gather*}
or, in the matrix form,
\begin{gather}
\mypmatrix{A&-B\cr-C&D}\mypmatrix{\tdhlshifted{X}{\tau}\cr\tdhlshifted{NX}{\tau}}=\mypmatrix{\mathfrak{d}
^{(1)}\cr\mathfrak{d}^{(2)}},
\label{app-matrix}
\end{gather}
where $\mathfrak{d}^{(1,2)}$ are linear combinations of shifted $\tdhlz{5}$ and $\tdhlz{6}$. 
On the other hand, one can straightforwardly verify the identity
\begin{gather}
\sigma\tdhlShifted{W}{\sigma}\tdhlZ{3}+\mu^{-1}(A D-B C)+\tdhlShifted{W}{\sigma}^{2}\tdhlz{0}+\sigma^{2}
\tdhlShifted{W}{\tdhlz{0}}=0.
\label{app-P}
\end{gather}
Equations \eqref{eqs-to-solve} imply vanishing of the last two terms of the left-hand side of the last
equation and of the vector $(\mathfrak{d}^{(1)},\mathfrak{d}^{(2)})^{T}$, which means that the determinant
of the matrix that appears in the left-hand side of \eqref{app-matrix} is zero, $AD-BC=0$, since we assume
$\tau\ne0$ (and, hence, $\tdhlshifted{X}{\tau}\ne0$ and $\tdhlshifted{NX}{\tau}\ne0$). 
This, together with \eqref{app-P}, leads to
\begin{gather*}
\tdhlz{0}=\tdhlz{5}=\tdhlz{6}=0
\qquad
\Rightarrow
\qquad
\tdhlZ{3}=0
\end{gather*}
which completes the proof of the fact that equations \eqref{alh-a}--\eqref{alh-e} imply \eqref{eqs-to-solve}
which, in its turn, completes the proof of Proposition~\ref{prop-alh}.

\section{Verif\/ication of the solitonic ansatz}
\label{app-ver}

To make the following formulae more readable we use here the `node' notation not only for the nodes of
lattices (direct and dual) but for the results of the auxiliary shifts $\tdhlshifted{X,Y}{}$ as well:
\begin{gather*}
f\mysub{X}=\tdhlshifted{X}{f},
\qquad
f\mysub{Y}=\tdhlshifted{Y}{f},
\qquad
f\mysub{SX}=\tdhlshifted{X}{f\mysub{S}}=\tdhlshifted{X}{\tdhlshifted{S}f},
\qquad
\text{{etc}}. 
\end{gather*}

First let us prove the fact that equations \eqref{sols-qr}--\eqref{sols-shifts-bra} imply \eqref{alh-b}.
Applying \eqref{rank-one} and the $S$-part of \eqref{sols-shift-SXY} to the product
$\tdhlmatrix[S]{B}\tdhlmatrix[S]{A}$ one can get
\begin{gather*}
\tdhlmatrix{B}\tdhlmatrix{A}-\tdhlmatrix[S]{B}\tdhlmatrix[S]{A}=\tdhlmatrix[S]{B}\tdhlket{1}\tdhlbra{a}
\end{gather*}
which leads to
\begin{gather}
\tdhlmatrix{F}\tdhlmatrix[S]{F}^{-1}=\tdhlmatrix{1}+\tdhlmatrix{F}\tdhlmatrix[S]{B}\tdhlket{1}\tdhlbra{a}
\label{sols-sa}
\end{gather}
and
\begin{gather*}
\tdhlmatrix[S]{F}\tdhlmatrix{F}^{-1}=\tdhlmatrix{1}-\tdhlmatrix[S]{F}\tdhlmatrix[S]{B}\tdhlket{1}\tdhlbra{a}.
\end{gather*}
Shifting this equation in the $\tdhlT{N}\tdhlT{X}$-direction one can obtain
\begin{gather}
\tdhlmatrix{K}\tdhlmatrix[X]{F}\tdhlmatrix[NX]{F}^{-1}\tdhlmatrix{K}^{-1}=\tdhlmatrix{1}-\tdhlmatrix{K}
\tdhlmatrix[X]{F}\tdhlmatrix[X]{B}\tdhlket{1}\tdhlbra{a},
\label{sols-xa}
\end{gather}
where
\begin{gather*}
\tdhlmatrix{K}=\tdhlR^{-1}+\mu
\end{gather*}
(we have used the fact that $\tdhlbra[NX]{a}\tdhlmatrix{K}^{-1}=\tdhlbra{a}$).
Multiplying \eqref{sols-sa} and \eqref{sols-xa} one arrives at
\begin{gather}
\tdhlmatrix{F}\tdhlmatrix[S]{F}^{-1}\;\tdhlmatrix{K}\tdhlmatrix[X]{F}\tdhlmatrix[NX]{F}^{-1}\tdhlmatrix{K}
^{-1}=\tdhlmatrix{1}+\tdhlmatrix{F}\tdhlmatrix{U}\tdhlket{1}\tdhlbra{a},
\label{sols-eq-U}
\end{gather}
where
\begin{gather*}
\tdhlmatrix{U}=\tdhlmatrix[S]{B}-\tdhlmatrix[S]{F}^{-1}\tdhlmatrix{K}\tdhlmatrix[X]{F}\tdhlmatrix[X]{B}
=\tdhlmatrix[S]{B}\left(\tdhlmatrix{1}+\tdhlmatrix[S]{A}\tdhlmatrix{K}\tdhlmatrix[X]{F}\tdhlmatrix[X]{B}
\right)-\tdhlmatrix{K}\tdhlmatrix[X]{F}\tdhlmatrix[X]{B}.
\end{gather*}
Noting that $\tdhlmatrix[S]{A}\tdhlmatrix{K}=\tdhlmatrix[X]{A}$ and applying the identities
\begin{gather*}
\tdhlmatrix{F}\tdhlmatrix{B}=\tdhlmatrix{B}\tdhlmatrix{G},
\qquad
\tdhlmatrix{1}+\tdhlmatrix{A}\tdhlmatrix{F}\tdhlmatrix{B}=\tdhlmatrix{G}
\end{gather*}
one can continue the calculations as follows:
\begin{gather*}
\tdhlmatrix{U}=\tdhlmatrix[S]{B}\left(\tdhlmatrix{1}+\tdhlmatrix[X]{A}\tdhlmatrix[X]{F}\tdhlmatrix[X]{B}
\right)-\tdhlmatrix{K}\tdhlmatrix[X]{F}\tdhlmatrix[X]{B}
=\left(\tdhlmatrix[S]{B}-\tdhlmatrix{K}\tdhlmatrix[X]{B}\right)\tdhlmatrix[X]{G}
=\tdhlR^{-1}\tdhlket{1}\tdhlbra[SX]{b}\tdhlmatrix[X]{G}.
\end{gather*}
After substitution of $\tdhlmatrix{U}$ in \eqref{sols-eq-U} and calculating the determinants of the both
sides one arrives~at
\begin{gather*}
\frac{\tau\mysub{S}\tau\mysub{NX}}{\tau\tau\mysub{X}}=1-q r\mysub{X}
\end{gather*}
(here, the identities $\tau=1/\det\tdhlmatrix{F}$ and
$\det(\tdhlmatrix{1}+|\mathsf{u}\rangle\langle\mathsf{v}|)=1+\langle\mathsf{v}|\mathsf{u}\rangle$ have been
used) which at the level of the tau-functions is
\begin{gather*}
0=\tau\tau\mysub{X}-\sigma\rho\mysub{X}-\tau\mysub{S}\tau\mysub{NX}.
\end{gather*}
Thus we have proved that our solutions satisfy equations \eqref{alh-b}.

Noting that the implementation of the shift $\tdhlT{X}$ depends analytically on $\mu$ and
$\left.\tdhlT{X}\right|_{\mu=0}$ is the unit operator, one obtains from the above calculations, by sending
$\mu\to0$, that our ansatz ensures \eqref{alh-a} as well.

In a~similar way, applying \eqref{rank-one} to $\tdhlmatrix[X]{B}\tdhlmatrix[X]{A}$ one can obtain
\begin{gather*}
\tdhlmatrix[X]{F}\tdhlmatrix{F}^{-1}=\tdhlmatrix{1}-\mu\tdhlmatrix[X]{F}\tdhlmatrix[X]{B}\tdhlket{1}
\tdhlbra{a},
\qquad
\tdhlmatrix{F}\tdhlmatrix[X]{F}^{-1}=\tdhlmatrix{1}+\mu\tdhlmatrix{F}\tdhlmatrix[X]{B}\tdhlket{1}
\tdhlbra{a}
\end{gather*}
and then
\begin{gather*}
\tdhlmatrix[S]{F}\tdhlmatrix[SX]{F}^{-1} \tdhlR^{-1}\tdhlmatrix[X]{F}\tdhlmatrix{F}^{-1}
\tdhlR=\tdhlmatrix{1}+\mu\tdhlmatrix[S]{F}\tdhlmatrix{V}\tdhlket{1}\tdhlbra[S]{a},
\end{gather*}
where
\begin{gather*}
\tdhlmatrix{V}=\tdhlmatrix[SX]{B}-\tdhlmatrix[SX]{F}^{-1}\tdhlR^{-1}\tdhlmatrix[X]{F}\tdhlmatrix[X]{B}
=\tdhlmatrix[SX]{B}\left(\tdhlmatrix{1}+\tdhlmatrix[X]{A}\tdhlmatrix[X]{F}\tdhlmatrix[X]{B}
\right)-\tdhlR^{-1}\tdhlmatrix[X]{F}\tdhlmatrix[X]{B}
\\
\phantom{\tdhlmatrix{V}}
=\left(\tdhlmatrix[SX]{B}-\tdhlR^{-1}\tdhlmatrix[X]{B}\right)\tdhlmatrix[X]{G}
=\tdhlR^{-1}\tdhlket{1}\tdhlbra[SX]{b}\tdhlmatrix[X]{G},
\end{gather*}
which leads to
\begin{gather*}
\frac{\tau\tau\mysub{SX}}{\tau\mysub{S}\tau\mysub{X}}=1-\mu q\mysub{S}r\mysub{X}
\end{gather*}
and
\begin{gather*}
0=\mu\sigma\mysub{S}\rho\mysub{X}-\tau\mysub{S}\tau\mysub{X}+\tau\tau\mysub{SX}.
\end{gather*}
This proves the fact that our solutions satisfy \eqref{alh-c}.

Considering the $\tdhlT{Y}$-equations, we do not present here the calculations similar to ones discussed
above, leaving the verif\/ication of \eqref{alh-d} and \eqref{alh-e} to the reader.

\section*{Acknowledgements}

We would like to thank the referees for careful reading the manuscript and for providing useful comments
and suggestions which helped us to improve the paper.

\pdfbookmark[1]{References}{ref}
\LastPageEnding

\end{document}